\documentclass[aps,prd,twocolumn]{revtex4-1}
\usepackage{epsfig,dsfont,amssymb,amsmath,amsthm,amsfonts,amsbsy,mathrsfs,mathtools,appendix}
\usepackage{graphicx}
\usepackage{float}
\usepackage{color}
\usepackage{cases}
\usepackage{scrtime}
\usepackage{rotating}
\usepackage{multirow} 
\usepackage{CJKutf8}
\usepackage{enumerate}

\newcommand\aap{Astron. Astrophys.}

\newcommand\apjl{Astrophys. J. Lett.}

\newcommand\epjc{Eur. Phys. J. C}

\newcommand\grg{Gen. Relativ. Gravit.}

\newcommand\ijmpd{Int. J. Mod. Phys. D}

\newcommand\jcap{J. Cosmol. Astropart. Phys.}

\newcommand\jmp{J. Math. Phys.}

\newcommand\lrr{Living Rev. Relativ.}
\newcommand\mnras{Mon. Not. R. Astron. Soc.}

\newcommand\physrep{Phys. Rep.}

\newcommand\sciadv{Sci. Adv.}

\begin{document}
\title{Analytical description for light propagation with the source inside the black hole photon sphere}

\author{Yuan-Xing Gao}
\email{gao$_$yx@outlook.com}
\affiliation{Purple Mountain Observatory, Chinese Academy of Sciences, Nanjing 210023, China
}

\begin{abstract}
The photon sphere defines the unstable circular orbit of photons in a black hole spacetime.
Photons emitted by a source located inside the photon sphere can be gravitationally lensed by the black hole and have time delays when reaching the observer. These delays may lead to light echoes produced in the light curve if an accretion event in the vicinity of the horizon can be observed. 
In this work, we present fully analytical formulas with high accuracy to describe the change of the azimuthal angle and the travel time of those photons.
By employing the analytical approaches, we find that the time delay between photons emitted from the interior of the photon sphere has a typical time scale of $2(\pi - \phi_\mathrm{S}) u_\mathrm{m}$ with $\phi_\mathrm{S}$ and $u_\mathrm{m}$ being respectively the azimuthal angle of the source and the impact parameter evaluated at the photon sphere, which can provide some clues on the future search for gravitational lensing signatures in the accretion inflow event.
\end{abstract}

\maketitle

\allowdisplaybreaks

\section{Introduction}
\label{sec_intro}

In the spacetime of a black hole, the deflection of photons can give rise to the phenomenon of gravitational lensing, which serves as a powerful tool for studying the physical properties of the black hole and has been extensively studied in the literature \cite{Perlick2004LRR7.9,Bozza2010GRG42.2269,Cunha2018GRG50.42}. 
In the asymptotically flat region far away from the black hole, photons get weakly deflected and two images may be generated due to gravitational lensing \cite{Keeton2005PRD72.104006}. These images have the potential to be confirmed by the monitoring of stars near a supermassive black hole by GRAVITY and to play a crucial role in understanding the structure of the spacetime \cite{Keeton2006PRD73.044024,GRAVITY2018A&A618.L10}. 
In the vicinity of the black hole event horizon, photons can be strongly deflected and gravitational lensing generates infinite relativistic images and photon rings that are important observational targets for the next-generation Event Horizon Telescope or future space-borne very long baseline interferometry \cite{EHTC2019ApJ875.L1,EHTC2022ApJ930.L12,Johnson2020SciAdv6.eaaz1310,Vincent2022AAp667.A170,Paugnat2022AAp668.A11}, enabling our in-depth exploration of the strong-field physics \cite{Gralla2019PRD100.024018,Wang2019JCAP04.022,Lu2019EPJC79.1016,Gao2021PRD103.043008,Zhang2022EPJC82.471,Aratore2021JCAP10.054,Gao2024PRD109.063030}.

The key element determining whether the photons deflected by the black hole can be observed is the photon sphere that defines the unstable circular orbit of photons \cite{Claudel2001JMP42.818}. 
For photons emitted by a source outside the photon sphere, their initial direction can be radially outward or inward.
The outward photons can be received if they are directly facing the observer.
Regarding those initially moving inward, they can reach the observer after being deflected only if their impact parameter (defined as the ratio of angular momentum to energy) is larger than that evaluated at the photon sphere; otherwise, they will be absorbed by the black hole \cite{Bozza2010GRG42.2269}. 
However, a different situation arises for photons emitted by a source inside the photon sphere.
Although both the initially inward-moving photons and the initially outward-moving photons with impact parameter larger than that at the photon sphere will be absorbed by the black hole,
the outward-moving photons with impact parameter smaller than that at the photon sphere can escape and reach the observer \cite{Bozza2007PRD76.083008}. The latter will be the main focus of this work.

Photons emitted by the source inside the photon sphere are crucial for forming the near-horizon image of a black hole. The profile of the so-called inner shadow offers a direct view of the event horizon \cite{Dokuchaev2018IJMPD28.1941005}, which requires less knowledge of accretion physics compared with current images of the external accretion disk surrounding the black hole and helps to break the degeneracies among the physical characteristics such as the mass, spin, and viewing angle of the black hole \cite{Chael2021ApJ918.6}. The polarization patterns of the near-horizon image are useful tools to study the frame-dragging effect, which usually occurs in the region inside the photon sphere \cite{Chen2024arXiv2407.14897,Hou2024arXiv2409.07248}. In addition, the strong deflection of photons in the accretion process may generate light echoes in the light curve \cite{CardenasAvendano2024PRL133.131402}. Since the accretion flow that enters the innermost stable circular orbit often reaches the event horizon quickly \cite{Cardoso2021PRD103.104044,Mummery2022PRL129.161101}, these echoes coming from the region inside the photon sphere may have a clean background and thus be possible to be extracted from the future interferometric signatures \cite{Wong2024APJL2024975.L40}.

To investigate the deflection of photons emitted by the source inside the photon sphere in the black hole spacetime, either numerical or analytical methods can be employed \cite{Perlick2022PR947.1}. 
Numerical methods typically involve forward or backward ray tracing algorithms, which require direct integration of the geodesic equations of photons \cite{Zhou2024arXiv2408.16049}. 
Alternatively, analytical approaches usually obtain approximate solutions for key physical quantities like the deflection angle and travel time through series expansion \cite{Keeton2005PRD72.104006,Bozza2002PRD66.103001,Bozza2004GRG36.435,Bozza2007PRD76.083008,Jia2021EPJC81.242,Liu2021EPJC81.894,Poutanen2020A&A640.A24,Claros2024PRD109.124055,Igata2025arXiv2503.02320}, and can reveal the underlying physics behind the practical observations more effectively.
Among all the analytical methods, the one based on the photons' impact parameter can describe the strong deflection of photons near the photon sphere in a completely analytical manner \cite{Jia2021EPJC81.242}, which has been successfully applied in the case where the source is outside the photon sphere. In this work, we will generalize this method to the case where the source is inside the photon sphere.

This paper is organized as follows. 
In Sec.~\ref{sec_spacetime_and_geodesic}, we introduce the general setup for the photon's propagation in a black hole spacetime when the source is inside the photon sphere.
In Sec.~\ref{sec_formulas_BendingAngel}, we present the fully analytical approximation for the bending angle, and evaluate the accuracy of our formulas.
In Sec.~\ref{sec_formulas_TravelTime}, we also give the fully analytical description for the travel time. 
From a practical perspective, we further apply the obtained formulas to calculate the time delay between two different photons emitted by the same source inside the photon sphere.
In Sec.~\ref{sec_conclusion}, we conclude and discuss our results.

\section{Light propagation with the source inside the photon sphere in the black hole spacetime}
\label{sec_spacetime_and_geodesic}

\subsection{Classification of different photons in the black hole spacetime}

A static spherically symmetric black hole in the asymptotically flat spacetime can be described by the following line element  
\begin{equation}\label{MetricBH}
\mathrm{d}s^2 = -A(r)\mathrm{d}t^2+B(r)\mathrm{d}r^2+C(r)(\mathrm{d}\theta^2+\sin^2\theta\mathrm{d}\phi^2),
\end{equation}
where $B(r)^{-1} = 0$ has at least one real root to ensure the existence of an event horizon.  

A photon with $\mathrm{d}s^2 = 0$ moves around the black hole along the geodesic as
\begin{equation}\label{geodesic}
    A(r) B(r) \dot{r}^2 + L^2\frac{A(r)}{C(r)} = E^2,
\end{equation}
where $\theta = \pi/2$ is fixed by taking advantage of the spherical symmetry. Here $E = A(r) \dot{t}$ and $L = C(r) \dot{\phi}$ are respectively the energy and angular momentum of that photon, and their ratio defines the photon's impact parameter
\begin{eqnarray}\label{def_u}
    u \equiv \frac{L}{E}.
\end{eqnarray}
Influenced by the black hole, the photon does not move along a straight line as it does in the flat spacetime, but rather is deflected. 

If the photon is emitted by a source far away from the black hole, the corresponding deflection angle is determined by the closest distance of approach (i.e., the turning point in the photon's trajectory).
The closer the closest distance to the black hole is, the larger the deflection angle will be. When the closest distance attains a critical value, the deflection angle diverges \cite{Claudel2001JMP42.818,Bozza2007PRD76.083008}. 
This critical value, denoted as $r_\mathrm{m}$, satisfies the following equation \cite{Virbhadra2000PRD62.084003,Claudel2001JMP42.818,Adler2022GRG54.93}
\begin{eqnarray}\label{def_ps}
    \frac{C'(r_\mathrm{m})}{C(r_\mathrm{m})} - \frac{A'(r_\mathrm{m})}{A(r_\mathrm{m})} = 0,
\end{eqnarray}
and defines a two-dimensional spherical sphere, which is referred to as the photon sphere.

When the closest distance of the photon is equal to the radius $r_\mathrm{m}$, its impact parameter can be calculated by
\begin{eqnarray}\label{def_um}
    u_\mathrm{m} = \sqrt{\frac{C_\mathrm{m}}{A_\mathrm{m}}},
\end{eqnarray}
where the subscript ``m" denotes functions evaluated at $r = r_\mathrm{m}$. 
Photons with $u = u_\mathrm{m}$ will wind around the black hole infinite times.

\begin{figure}[htbp]    
	\centering
	\includegraphics[width=0.4\textwidth]{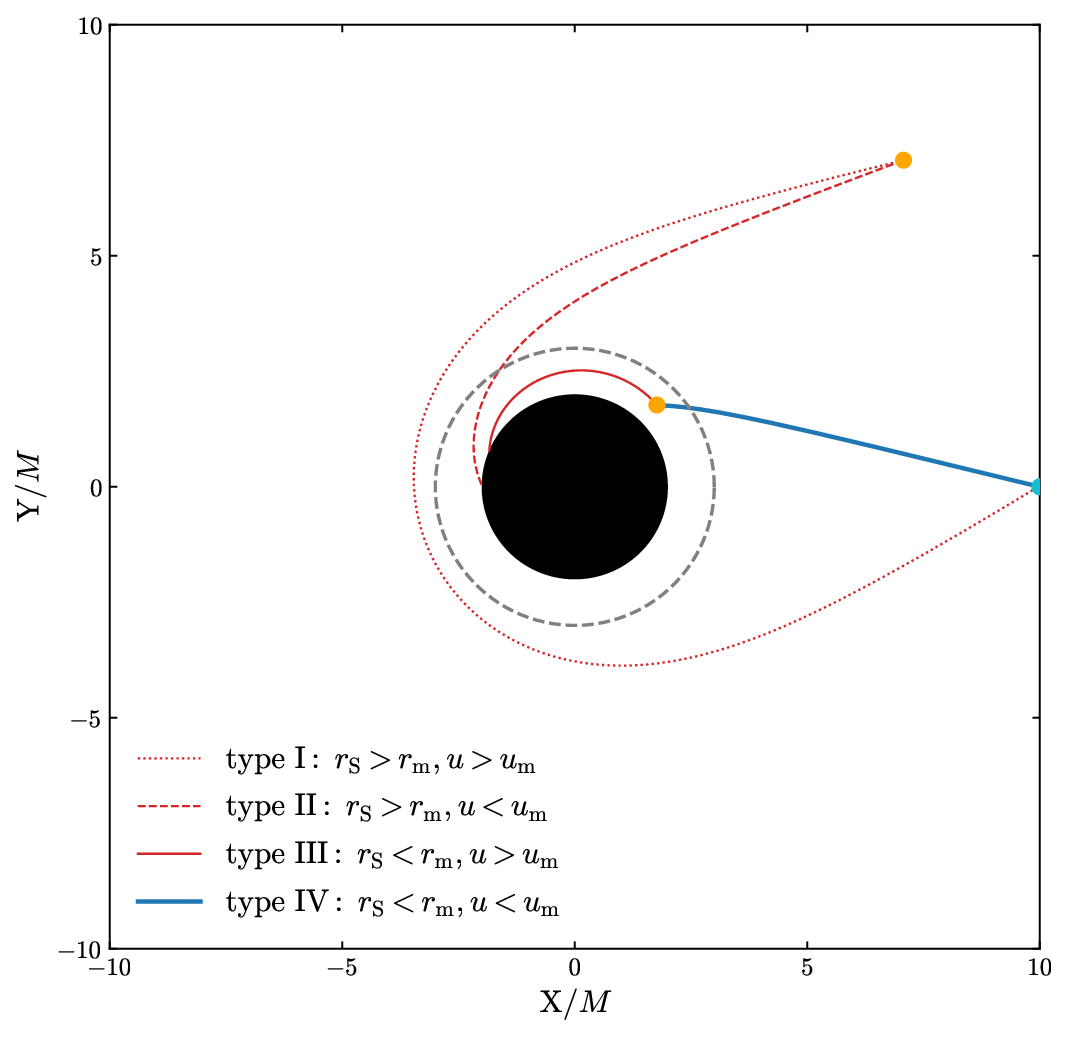}
	\caption{Different types of photons in a black hole spacetime. The central dark region is bounded by the black hole event horizon, the orange circles represent two sources with different locations, the sky blue circle represents the observer located at $(10 M, 0)$, and the dashed grey circle represents the photon sphere.}
	\label{Fig1_Photon_Classification_of_BH}
\end{figure}

As shown in Fig.~\ref{Fig1_Photon_Classification_of_BH}, according to the impact parameter $u$ of photons and the position $r_\mathrm{S}$ of the source, we can classify the photons in a black hole spacetime into the following four types.
\begin{itemize}
    \item[(\uppercase\expandafter{\romannumeral 1})] Photons with $u > u_\mathrm{m}$ and $r_\mathrm{S} > r_\mathrm{m}$. 
    As shown by the red dotted line, these photons are emitted by the source outside the photon sphere, and their initial direction can be either radially inward or outward. 
    For the inward ones, they can be received by the observer after being weakly deflected when $u$ is much larger than $u_\mathrm{m}$ \cite{Keeton2005PRD72.104006,Huang2020JCAP08.016} or being strongly deflected when $u$ is slightly larger than $u_\mathrm{m}$ \cite{Claudel2001JMP42.818,Bozza2002PRD66.103001,Jia2021EPJC81.242}. For the outward ones, they can also be observed if they are directly facing the observer.
    \item[(\uppercase\expandafter{\romannumeral 2})] Photons with $u < u_\mathrm{m}$ and $r_\mathrm{S} > r_\mathrm{m}$.
    They also come from the region outside the photon sphere, as indicated by the red dashed line.  
    Although photons having an initially radially-outward direction might still be detectable, those whose initial direction is radially-inward will enter the photon sphere and be absorbed by the black hole due to the condition $u < u_\mathrm{m}$ \cite{Gralla2019PRD100.024018}.
    \item[(\uppercase\expandafter{\romannumeral 3})] Photons with $u > u_\mathrm{m}$ and $r_\mathrm{S} < r_\mathrm{m}$.
    They originate from the region inside the photon sphere and outside the event horizon. Given that $u > u_\mathrm{m}$, these photons feature turning points in their trajectories and will eventually fall into the horizon \cite{Claudel2001JMP42.818,Bozza2007PRD76.083008}, as depicted by the red solid line. 
    \item[(\uppercase\expandafter{\romannumeral 4})] Photons with $u < u_\mathrm{m}$ and $r_\mathrm{S} < r_\mathrm{m}$. 
    These photons are emitted by the source inside the photon sphere and are very close to the event horizon. If their initial direction is radially outward, they can escape and be received by the observer \cite{Bozza2007PRD76.083008}, as illustrated by the blue solid line. 
    This type of photons will be our focus.
\end{itemize}

For the type \uppercase\expandafter{\romannumeral 4} photons that are characterized by $u < u_\mathrm{m}$ and $r_\mathrm{S} < r_\mathrm{m}$, 
they can be either weakly deflected or strongly deflected before being received by the observer, as in the case where the source is located outside the photon sphere. 
This fact has been confirmed numerically \cite{Gralla2019PRD100.024018}. 
Ref.~\cite{Bozza2007PRD76.083008} presented an analytical approach to describe these photons, but its accuracy for the portion of weak deflection is insufficient; the methods proposed in Refs.~\cite{Poutanen2020A&A640.A24,Claros2024PRD109.124055} can match the weak deflection portion very well, but need to be generalized to cover the portion of strong deflection.
To address these issues, we shall present fully analytical formulas to accurately describe all portions of the type \uppercase\expandafter{\romannumeral 4} photons.

\subsection{Light propagation with the source inside the photon sphere}

For photons of type \uppercase\expandafter{\romannumeral 4} propagating in the black hole spacetime, 
the change in the azimuthal angle and the travel time can be respectively obtained as \cite{Virbhadra1998AA337.1}
\begin{eqnarray}\label{DeltaPhi_InBHPS}
\Delta\phi &=& \int_{r_\mathrm{S}}^{r_\mathrm{O}}\frac{\dot{\phi}}{\dot{r}}\mathrm{d}r \nonumber\\
&=& \int_{r_\mathrm{S}}^{r_\mathrm{O}} \sqrt{\frac{u^2A(r)B(r)}{C(r)[C(r) - u^2A(r)]}}\ \mathrm{d}r,
\end{eqnarray}
and \cite{Virbhadra2008PRD77.124014}
\begin{eqnarray}\label{T_InBHPS}
T &=& \int_{r_\mathrm{S}}^{r_\mathrm{O}}\frac{\dot{t}}{\dot{r}}\mathrm{d}r \nonumber\\
&=& \int_{r_\mathrm{S}}^{r_\mathrm{O}} \sqrt{\frac{B(r)C(r)}{A(r)[C(r) - u^2A(r)]}}\ \mathrm{d}r,
\end{eqnarray}
where $r_\mathrm{S}$ and $r_\mathrm{O}$ are the distance of the lens to the source and to the observer, respectively. 
As stated before, there are no turning points in the trajectories of those photons \cite{Bozza2007PRD76.083008}.
$\Delta\phi$ and $T$ are two key quantities for the light propagation: $\Delta\phi$ connects the geometric positions of the source, lens, and observer through the lens equation, and $T$ is closely related to the characteristics of the signal received by the observer.

To analytically solve the integrals \eqref{DeltaPhi_InBHPS} and \eqref{T_InBHPS}, we can change the variable of integration $r$ to $\xi$ via \cite{Jia2021EPJC81.242}
\begin{eqnarray}\label{e_and_xi_def}
    \xi = 1 - \frac{u_\mathrm{m}}{\sqrt{\frac{C(r)}{A(r)}}}.
\end{eqnarray}
According to the above definition, we can know that the transformation from $r$ to $\xi$ does not always have an explicit form in arbitrary spacetimes. 
Given that we aim to replace $r$ with $\xi$ in the integrand of \eqref{DeltaPhi_InBHPS} or \eqref{T_InBHPS}, it is necessary to introduce the following inverse function \cite{Jia2021EPJC81.242}
\begin{eqnarray}\label{q_def}
    q ( \zeta ) = \frac{1}{r},
\end{eqnarray}
where
\begin{eqnarray}\label{zeta_def}
    \zeta = \frac{\xi}{u_\mathrm{m}}.
\end{eqnarray}
For later use, we also denote the inverse function of $q$ as $p$ with \cite{Jia2021EPJC81.242}
\begin{eqnarray}\label{p_def}
    p\bigg( \frac{1}{r} \bigg) = \zeta = \frac{1}{u_\mathrm{m}} - \frac{1}{\sqrt{\frac{C(r)}{A(r)}}}.
\end{eqnarray}

Then the change in the azimuthal angle \eqref{DeltaPhi_InBHPS} can be rewritten as 
\begin{eqnarray}\label{DeltaPhi_InBHPS_new}
\Delta\phi &=& \int_{\eta_\mathrm{S}}^{\eta_\mathrm{O}}\frac{\bar{y}(\xi, \epsilon)}{\sqrt{\xi - \epsilon}}\mathrm{d}\xi,
\end{eqnarray}
where
\begin{eqnarray}\label{epsilon_def}
    \epsilon = 1 - \frac{u_\mathrm{m}}{u},
\end{eqnarray}
and
\begin{eqnarray}\label{ybar_def}
\bar{y}(\xi, \epsilon) &=& \frac{\bar{f}(\xi)}{\sqrt{2-\epsilon-\xi}}, \\
\bar{f}(\xi) &=& \sqrt{\frac{B(1/q)}{C(1/q)}} \frac{\xi - 1}{u_\mathrm{m}q^2} \frac{\mathrm{d}q}{\mathrm{d}\zeta}, \\
\eta_{S} &=& 1 - \frac{u_\mathrm{m}}{\sqrt{\frac{C(r_{S})}{A(r_{S})}}},\ \ \eta_{O} = 1 - \frac{u_\mathrm{m}}{\sqrt{\frac{C(r_{O})}{A(r_O)}}}.
\end{eqnarray}
Note that the parameter $\epsilon$ was previously used to calculate the bending angle in the Schawarzschild black hole spacetime \cite{Iyer2007GRG39.1563}. 
For photons emitted by a source inside the photon sphere, we have $\epsilon<0$.
Moreover, we find that $ 0 \leq \xi \leq 1$ for any position outside the event horzion.

Similarly, the travel time \eqref{T_InBHPS} expressed in terms of $\xi$ is given by
\begin{eqnarray}\label{T_InBHPS_new}
T &=& \int_{\eta_\mathrm{S}}^{\eta_\mathrm{O}}\frac{\tilde{y}(\xi, \epsilon)}{\sqrt{\xi - \epsilon}}\mathrm{d}\xi,
\end{eqnarray}
where
\begin{eqnarray}
\tilde{y}(\xi, \epsilon) &=& \frac{\tilde{f}(\xi)}{\sqrt{2-\epsilon-\xi}}, \\
\tilde{f}(\xi) &=& \frac{\sqrt{B(1/q)C(1/q)}}{A(1/q)} \frac{(\xi - 1)(1-\epsilon)}{u_\mathrm{m}^2q^2} \frac{\mathrm{d}q}{\mathrm{d}\zeta}.
\end{eqnarray}

The Integrals \eqref{DeltaPhi_InBHPS_new} and \eqref{T_InBHPS_new} now become the targets that we are going to solve analytically. 
We will carry them out in Sec.~\ref{sec_formulas_BendingAngel} and \ref{sec_formulas_TravelTime}, respectively.

\section{Analytical calculation of the bending angle with the source inside the photon sphere}
\label{sec_formulas_BendingAngel}

At the first step of solving the integral \eqref{DeltaPhi_InBHPS_new}, we need to convert its integrand into an integrable form.

In the integrand of \eqref{DeltaPhi_InBHPS_new}, $\bar{y}(\xi, \epsilon)$ is the product of $\bar{f}(\xi)$ and $(2 - \epsilon - \xi)^{-1/2}$, where $(2 - \epsilon - \xi)^{-1/2}$ satisfies \cite{Jia2021EPJC81.242}
\begin{eqnarray}\label{sqrt_for_DeltaPhi}
    \frac{1}{\sqrt{2 - \epsilon - \xi}} = \sum_{n=0}^\infty \frac{(2n - 1)!!}{(2n)!!}\frac{\xi^n}{(2-\epsilon)^{n+\frac{1}{2}}}.
\end{eqnarray}
Since the deflection of photons mainly occurs in the vicinity of the photon sphere, we can treat $\xi$ as the small quantity and expand $\bar{f}(\xi)$ as
\begin{eqnarray}\label{fxi_for_DeltaPhi}
\bar{f}(\xi) &=& \left\{ 
        \begin{aligned} & \sum_{j = -1}^\infty \bar{f}_{j}^+ \xi^{\frac{j}{2}}, & \mathrm{for}\ r > r_\mathrm{m} \\
        & \sum_{j = -1}^\infty \bar{f}_{j}^- \xi^{\frac{j}{2}}, & \mathrm{for}\ r < r_\mathrm{m}
        \end{aligned}
        \right. ,
\end{eqnarray} 
in which the coefficients $\bar{f}_{j}^\pm$ obey the following relation
\begin{eqnarray}\label{fn_relation}
    \bar{f}_{2k}^+ = \bar{f}_{2k}^-,\, \, \bar{f}_{2k-1}^+ = - \bar{f}_{2k-1}^-,\, \, k = 0,1,2,\cdots.
\end{eqnarray}
The key to obtaining Eq.~\eqref{fxi_for_DeltaPhi} is to find the series expansion of $q$ with respect to $\xi$, whereas the details are given in Appendix.~\ref{App_Derivation_fxi}.

It should be noted that when the source is inside the photon sphere, the expansions in the cases of $r < r_\mathrm{m}$ and $r > r_\mathrm{m}$ need to be handled separately because the power of $\xi$ is $j/2$. 
In contrast, when the source is outside the photon sphere, a unified treatment is sufficient without the need of such separate handling \cite{Jia2021EPJC81.242}.

After combining Eq.~\eqref{fxi_for_DeltaPhi} with \eqref{sqrt_for_DeltaPhi}, we collect the power of $\xi$ and obtain the following form of Eq.~\eqref{ybar_def} 
\begin{eqnarray}\label{ybar_explicit}
    \bar{y}^{+(-)} = \sum_{n=-1}^\infty \bigg( \sum_{m=0}^{[\frac{n+1}{2}]}\frac{\epsilon^m}{(2-\epsilon)^{\big[\frac{n+1}{2}\big]+\frac{1}{2}}} \bar{y}_{n,m}^{+(-)} \bigg)\xi^{\frac{n}{2}},
\end{eqnarray}
where $[\frac{n+1}{2}]$ represents the largest integer not exceeding $(n+1)/2$. $\bar{y}_{n,m}^{+(-)}$ is the coefficient of $\epsilon^m$, which is given by 
\begin{eqnarray}\label{y_nm_def}
    \bar{y}_{n,m}^{+(-)} &=& \mathrm{Coefficient} \bigg\{ \sum_{m=0}^{[\frac{n+1}{2}]}\sum_{i=0}^{[\frac{n+1}{2}]-m}\frac{(2m-1)!!}{(2m)!!}\bar{f}_{n-2m}^{+(-)} \nonumber \\
    && \times C_{[\frac{n+1}{2}]-m}^i 2^{[\frac{n+1}{2}]-m-i} (-1)^i\epsilon^i, \epsilon^m \bigg\}.
\end{eqnarray}
Here $C_{n}^m = \frac{(n)!}{(m)!(n-m)!}$ denotes the binomial coefficient, and $\mathrm{Coefficient} \bigg\{ \mathrm{``Expression"}, \epsilon^m \bigg\}$ gives the coefficient of $\epsilon^m$ in ``Expression".
Given the relation \eqref{fn_relation}, we can have
\begin{eqnarray}\label{y_nm_relation}
    \bar{y}_{2k,m}^{+} = \bar{y}_{2k,m}^{-}, \ \ \bar{y}_{2k-1,m}^{+} = - \bar{y}_{2k-1,m}^{-}.
\end{eqnarray}

By substituting Eq.~\eqref{ybar_explicit} into \eqref{DeltaPhi_InBHPS_new}, we can write $\Delta\phi$ as 
\begin{eqnarray}\label{DeltaPhi_InBHPS_integrable}
    \Delta\phi &=& \sum_{n=-1}^\infty \sum_{m=0}^{[\frac{n+1}{2}]}\frac{\epsilon^m}{(2-\epsilon)^{\big[\frac{n+1}{2}\big]+\frac{1}{2}}} \bigg( \bar{y}_{n,m}^{-} \int_{\eta_\mathrm{S}}^{\eta_\mathrm{m}}\frac{\xi^{\frac{n}{2}}}{\sqrt{\xi - \epsilon}}\mathrm{d}\xi \nonumber\\
    && +\, \bar{y}_{n,m}^{+}\int_{\eta_\mathrm{m}}^{\eta_\mathrm{O}}\frac{\xi^{\frac{n}{2}}}{\sqrt{\xi - \epsilon}}\mathrm{d}\xi \bigg),
\end{eqnarray}
where $\eta_\mathrm{m} = 1 - u_\mathrm{m} / \sqrt{C(r_\mathrm{m}) / A(r_\mathrm{m})} = 0$ and $0 \leq \eta_\mathrm{S}, \eta_\mathrm{O} \leq 1$.
Since the integral $\int \xi^{\frac{n}{2}} / \sqrt{\xi - \epsilon}\ \mathrm{d}\xi$ can be always carried out (the formulas are given in Appendix.~\ref{App_integral_formulas}), Eq.~\eqref{DeltaPhi_InBHPS_integrable} finally gives
\begin{eqnarray}\label{Ana_app_for_dphi_in_PS}
    \Delta \phi &=& \sum_{k=0}^\infty \sum_{m=0}^{k}\frac{\epsilon^m}{(2-\epsilon)^{k+\frac{1}{2}}} \bigg\{ \bar{y}_{2k-1,m}^{+} \frac{\epsilon^k C_{2k}^k}{4^k} \bigg[ - 2\ln(-\epsilon) \nonumber\\
    && + 2\ln(\sqrt{\eta_\mathrm{O}} + \sqrt{\eta_\mathrm{O} - \epsilon}) + 2\ln(\sqrt{\eta_\mathrm{S}} + \sqrt{\eta_\mathrm{S} - \epsilon}) \nonumber\\
    && + \sum_{j=1}^k \frac{1}{jC_{2j}^j}\frac{4^j}{\epsilon^j} \bigg(\eta_\mathrm{S}^{j} \sqrt{1 -  \frac{\epsilon}{\eta_\mathrm{S}}} + \eta_\mathrm{O}^{j} \sqrt{1 -  \frac{\epsilon}{\eta_\mathrm{O}}} \bigg)  \bigg] \nonumber\\
    && + \bar{y}_{2k,m}^{+} \sum_{j=0}^k \frac{2C_k^j \epsilon^{k-j}}{2j+1} \bigg[ (\eta_\mathrm{O} - \epsilon)^{j+\frac{1}{2}} \nonumber\\
    && - (\eta_\mathrm{S} - \epsilon)^{j+\frac{1}{2}} \bigg] \bigg\},
\end{eqnarray}
which is applicable for photons emitted by the source inside the photon sphere with $u \in (0, u_\mathrm{m})$ and $r_\mathrm{S} \in (r_\mathrm{h}, r_\mathrm{m}]$, where $r_\mathrm{h}$ denotes the event horizon.

At the lowest order, Eq.~\eqref{Ana_app_for_dphi_in_PS} reduces to
\begin{eqnarray}\label{Ana_app_for_dphi_in_PS_order0}
    \Delta \phi &=& -\sqrt{2} \bar{y}_{-1,0}^{+} \ln(-\epsilon) + \sqrt{2} \bigg[ \bar{y}_{-1,0}^{+} \ln \big( 4 \sqrt{ \eta_\mathrm{O}\eta_\mathrm{S} } \big) \nonumber\\
    && + \bar{y}_{0,0}^{+} \big( \sqrt{ \eta_\mathrm{O} } - \sqrt{ \eta_\mathrm{S} } \big) \bigg] + \mathcal{O}(\epsilon),
\end{eqnarray}
where
\begin{eqnarray}
    \bar{y}_{-1,0}^{+} = \bar{f}_{-1}^+,\, \, \bar{y}_{0,0}^{+} = \bar{f}_{0}^+,
\end{eqnarray}
and 
\begin{eqnarray}
    \bar{f}_{-1}^+ &=& \sqrt{\frac{A_\mathrm{m}B_\mathrm{m}}{ A_\mathrm{m}C''_\mathrm{m} - A''_\mathrm{m}C_\mathrm{m} }},\\
    \bar{f}_{0}^+ &=& \frac{ A_\mathrm{m} }{ 3\sqrt{B_\mathrm{m}C_\mathrm{m}}(A_\mathrm{m}C''_\mathrm{m} - A''_\mathrm{m}C_\mathrm{m})^2 } \bigg\{ 3A_\mathrm{m}B_\mathrm{m}C'_\mathrm{m}C''_\mathrm{m} \nonumber\\
    && + C_\mathrm{m}^2 \big( 2A'''_\mathrm{m}B_\mathrm{m} - 3A''_\mathrm{m}B'_\mathrm{m} \big) + C_\mathrm{m} \big[ 3A_\mathrm{m}B'_\mathrm{m}C''_\mathrm{m} \nonumber\\
    && - B_\mathrm{m} \big( 3A''_\mathrm{m}C'_\mathrm{m} + 2A_\mathrm{m}C'''_\mathrm{m} \big) \big] \bigg\}.
\end{eqnarray}
According to Eq.~\eqref{Ana_app_for_dphi_in_PS_order0}, we can quickly evaluate the magnitude of $\Delta \phi$.

\subsection{Comparison with previous works and numerical results}
\label{subsec_comparison_DeltaPhi}

\subsubsection{Comparison with the result of Ref.~\cite{Jia2021EPJC81.242}}

The fully analytical approximation of $\Delta \phi$ for photons emitted by the source outside the photon sphere is found to be \cite{Jia2021EPJC81.242}
\begin{eqnarray}\label{Ana_app_for_dphi_out_PS}
    \Delta \phi_\mathrm{ops} &=& \sum_{k=0}^\infty \sum_{m=0}^{k}\frac{\epsilon^m}{(2-\epsilon)^{k+\frac{1}{2}}} \bigg\{ \bar{y}_{2k-1,m}^{+} \frac{\epsilon^k C_{2k}^k}{4^k} \bigg[ - 2\ln(\epsilon) \nonumber\\
    && + 2\ln(\sqrt{\eta_\mathrm{O}} + \sqrt{\eta_\mathrm{O} - \epsilon}) + 2\ln(\sqrt{\eta_\mathrm{S}} + \sqrt{\eta_\mathrm{S} - \epsilon}) \nonumber\\
    && + \sum_{j=1}^k \frac{1}{jC_{2j}^j}\frac{4^j}{\epsilon^j} \bigg(\eta_\mathrm{S}^{j} \sqrt{1 -  \frac{\epsilon}{\eta_\mathrm{S}}} + \eta_\mathrm{O}^{j} \sqrt{1 -  \frac{\epsilon}{\eta_\mathrm{O}}} \bigg)  \bigg] \nonumber\\
    && + \bar{y}_{2k,m}^{+} \sum_{j=0}^k \frac{2C_k^j \epsilon^{k-j}}{2j+1} \bigg[ (\eta_\mathrm{O} - \epsilon)^{j+\frac{1}{2}} \nonumber\\
    && + (\eta_\mathrm{S} - \epsilon)^{j+\frac{1}{2}} \bigg] \bigg\},
\end{eqnarray}
where the subscript ``ops" represents ``outside the photon sphere".

By comparing the above equation with our result \eqref{Ana_app_for_dphi_in_PS}, we can find that these two results are very similar, and the only differences between them are the logarithmic term and the sign of the last term. 
Eqs.\eqref{Ana_app_for_dphi_in_PS} and \eqref{Ana_app_for_dphi_out_PS} do not connect smoothly at the photon sphere and therefore cannot be unified.
The reasons are as follows.
For photons emitted by the source outside the photon sphere, we have $\epsilon \leq \eta_{i (i = \mathrm{O,S)} }$, which reveals that $\sqrt{ \eta_{i (i = \mathrm{O,S)} } - \epsilon } $ is always positive. 
However, for photons emitted by the source inside the photon sphere, the absolute value of $\epsilon (< 0)$ is not smaller than that of $ \eta_{i (i = \mathrm{O,S)} } (> 0)$ everywhere. 
If we just simply apply Eq.~\eqref{Ana_app_for_dphi_out_PS} to the case where the source is inside the photon sphere by replacing $\epsilon$ with $-\epsilon$, the value of $\sqrt{ \eta_{i (i = \mathrm{O,S)} } + \epsilon } $ may be imaginary and the final results may be problematic.

\subsubsection{Comparison with the result of Ref.~\cite{Bozza2007PRD76.083008}}

The leading-order analytical approximation of $\Delta \phi$ for photons emitted by the source inside the photon sphere can be given by \cite{Bozza2007PRD76.083008}
\begin{eqnarray}\label{Ana_app_for_dphi_leading_order}
    \Delta \phi & = & -\, a \ln \frac{(u_\mathrm{m} / u - 1)}{ ( 1 - r_\mathrm{m} / r_\mathrm{S} ) ( 1 - r_\mathrm{m} / r_\mathrm{O} ) } + b + \pi \nonumber\\
    & & +\, \mathcal{O} [ (u - u_\mathrm{m}) \ln(u - u_\mathrm{m}) ] ,
\end{eqnarray}
where $a$ and $b$ are the strong deflection coefficients that are directly associated with the metric functions $A(r_\mathrm{m}), B(r_\mathrm{m})$ and $C(r_\mathrm{m})$. Note that the value of $b$ usually needs to be obtained through numerical integration.

Eq.~\eqref{Ana_app_for_dphi_leading_order} has been demonstrated to be applicable to photons emitted by sources at any distance outside the event horizon \cite{Bozza2007PRD76.083008}. 
However, its accuracy is compromised when $u$ is considerably smaller than $u_\mathrm{m}$, and this is indeed the situation that we will take into consideration later.
In comparison, Eq.~\eqref{Ana_app_for_dphi_in_PS} is only suitable for the case where the source is inside the photon sphere, but gives the higher-order corrections that are omitted by Ref.~\cite{Bozza2007PRD76.083008}.  Therefore, our approximation can be more accurate.

\subsubsection{Comparison with numerical results}

\begin{figure*}[t!]    
	\centering
	\includegraphics[width=1.0\textwidth]{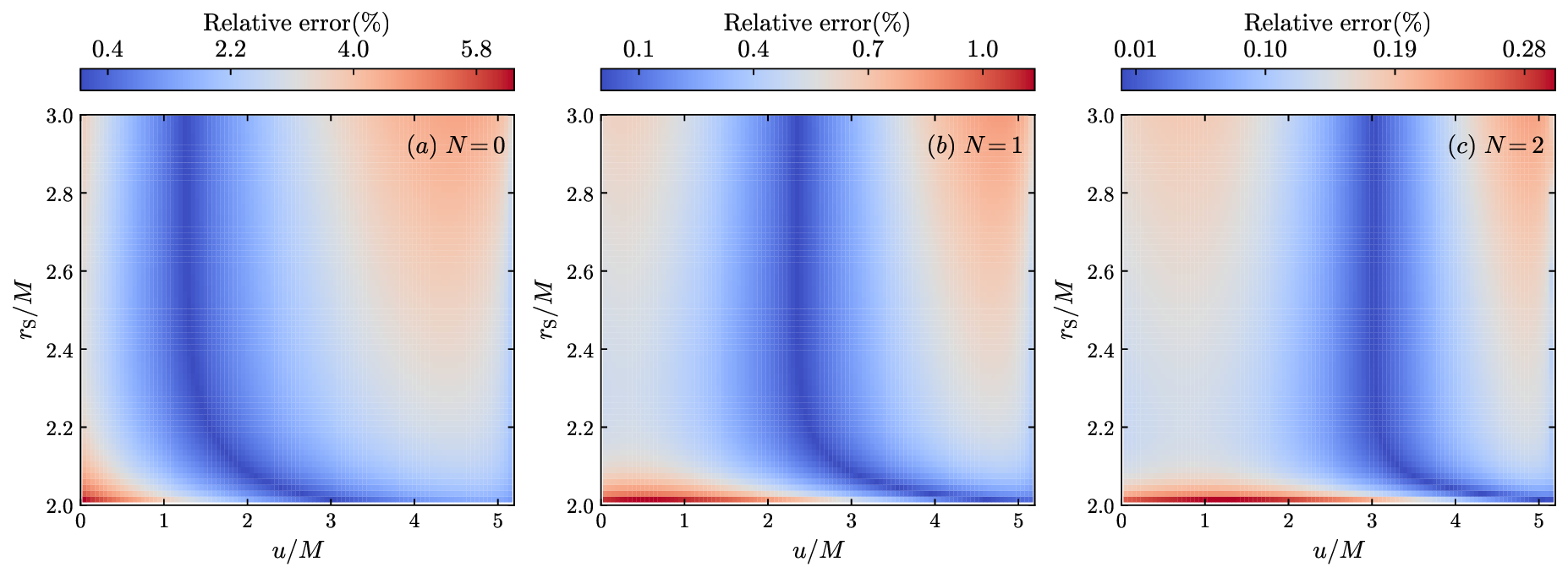}
	\caption{Relative errors between the analytical approximation \eqref{Ana_app_for_dphi_in_PS} and the numerical results in the Schwarzschild metric. The location of the source $r_\mathrm{S}$ varies from $2M$ to $3M$, and the impact parameter $u$ varies from $0$ to $3\sqrt{3}M$. The orders of approximation are respectively (a)$N = 0$; (b)$N = 1$; (c)$N = 2$. }
	\label{Fig2_RE_for_DeltaPhi}
\end{figure*}

In order to validate the analytical approximation \eqref{Ana_app_for_dphi_in_PS}, we need to set a truncation order $N$ by replacing $\sum_{k=0}^\infty$ with $\sum_{k=0}^N$ in Eq.~\eqref{Ana_app_for_dphi_in_PS}.
Assuming the spacetime is described by the following Schwarzschild metric $(G = c = M = 1)$
\begin{eqnarray}\label{MetricBH_Sch}
    A(r) & = & 1 - \frac{2}{r}, \\
    B(r) & = & \bigg( 1 - \frac{2}{r} \bigg)^{-1}, \\
    C(r) & = & r^2, 
\end{eqnarray}
we show the relative errors in Fig.~\ref{Fig2_RE_for_DeltaPhi} by evaluating $\big|( \Delta \phi - \Delta \phi_\mathrm{num} ) / \Delta \phi_\mathrm{num} \big|$, where $\Delta \phi_\mathrm{num}$ is the numerical result by employing the Simpson integration algorithm and setting the location of the observer $r_\mathrm{O} = 1000M$. The 
location of the source $r_\mathrm{S}$ varies from $2M$ to $3M$, and the impact parameter $u$ varies from $0$ to $3\sqrt{3}M$. From the first column to the third column, the truncation order of the approximation is $N = 0, 1, 2$, respectively.

We find that the lowest truncation order $(N = 0)$ that is given by Eq.~\eqref{Ana_app_for_dphi_in_PS_order0} has the maximum relative error of about $6\%$ for those photons emitted near the event horizon $(r_\mathrm{S} \approx 2)$ in a radial direction towards the observer $(u \approx 0)$.
As the truncation order $N$ increases, the relative error decreases significantly.  For both cases of $N = 1$ and  $N = 2$, the relative error is well below $1\%$, demonstrating the validity and high accuracy of Eq.~\eqref{Ana_app_for_dphi_in_PS}.

\section{Analytical calculation of the travel time}
\label{sec_formulas_TravelTime}

For photons emitted by the source inside the photon sphere, by adopting a similar methodology, we can obtain the analytical formulas for the travel time \eqref{T_InBHPS} as
\begin{eqnarray}\label{Ana_app_for_T_in_PS}
    T &=& \sum_{k=0}^\infty \sum_{m=0}^{k}\frac{(1 - \epsilon )\epsilon^m}{(2-\epsilon)^{k+\frac{1}{2}}} \bigg\{  \tilde{y}_{2k-1,m}^{+} \frac{\epsilon^k C_{2k}^k}{4^k} \bigg[ - 2\ln(-\epsilon) \nonumber\\
    && + 2\ln(\sqrt{\eta_\mathrm{O}} + \sqrt{\eta_\mathrm{O} - \epsilon}) + 2\ln(\sqrt{\eta_\mathrm{S}} + \sqrt{\eta_\mathrm{S} - \epsilon}) \nonumber\\
    && + \sum_{j=1}^k \frac{1}{jC_{2j}^j}\frac{4^j}{\epsilon^j} \bigg(\eta_\mathrm{S}^{j} \sqrt{1 -  \frac{\epsilon}{\eta_\mathrm{S}}} + \eta_\mathrm{O}^{j} \sqrt{1 -  \frac{\epsilon}{\eta_\mathrm{O}}} \bigg)  \bigg] \nonumber\\
    && +\, \tilde{y}_{2k,m}^{+} \sum_{j=0}^k \frac{2C_k^j \epsilon^{k-j}}{2j+1} \bigg[ (\eta_\mathrm{O} - \epsilon)^{j+\frac{1}{2}} \nonumber\\
    && - (\eta_\mathrm{S} - \epsilon)^{j+\frac{1}{2}} \bigg] \bigg\}.
\end{eqnarray}
At the lowest order, we can have
\begin{eqnarray}\label{Ana_app_for_T_in_PS_order0}
    T &=& - \sqrt{2}\tilde{y}_{-1,0}^{+} \ln(-\epsilon) + \sqrt{2} \bigg[\tilde{y}_{-1,0}^{+} \ln( 4\sqrt{\eta_\mathrm{O}\eta_\mathrm{S}} ) \nonumber\\
    && + \tilde{y}_{0,0}^{+} \big( \sqrt{\eta_\mathrm{O}} - \sqrt{\eta_\mathrm{S}} \big) \bigg] + \mathcal{O}(\epsilon),
\end{eqnarray}
where
\begin{eqnarray}
    \tilde{y}_{-1,0}^{+} &=& \tilde{f}_{-1}^+,\\
    \tilde{y}_{0,0}^{+} &=& \tilde{f}_{0}^+,
\end{eqnarray}
and
\begin{eqnarray}\label{relation_tf_and_bf}
    \tilde{f}_{-1}^+ = \bar{f}_{-1}^+ u_\mathrm{m},\, \, 
    \tilde{f}_{0}^+ = \bar{f}_{0}^+ u_\mathrm{m}.
\end{eqnarray}
It can be checked that Eq.~\eqref{Ana_app_for_T_in_PS} is convergent, whereas the actual travel time tends to diverge if the observer is far enough from the black hole. Consequently, the approximation \eqref{Ana_app_for_T_in_PS} is only applicable to describing the motion of photons in the near region of the black hole or calculating the time delay between two photons received by taking advantage of the fact that the integrand of \eqref{T_InBHPS} approaches unity and the resulting travel times of these photons can approximately cancel each other out for large $r$ \cite{Bozza2004GRG36.435}.

\subsection{Comparison with previous works and numerical results }
\label{subsec_comparison_T}

\subsubsection{Comparison with the result of Ref.~\cite{Liu2021EPJC81.894}}

The travel time of a photon emitted by the source outside the photon sphere is found to be \cite{Liu2021EPJC81.894}
\begin{eqnarray}\label{Ana_app_for_T_out_PS}
    T_\mathrm{ops} &=& \sum_{k=0}^\infty \sum_{m=0}^{k}\frac{(1 - \epsilon )\epsilon^m}{(2-\epsilon)^{k+\frac{1}{2}}} \bigg\{  \tilde{y}_{2k-1,m}^{+} \frac{\epsilon^k C_{2k}^k}{4^k} \bigg[ - 2\ln(\epsilon) \nonumber\\
    && + 2\ln(\sqrt{\eta_\mathrm{O}} + \sqrt{\eta_\mathrm{O} - \epsilon}) + 2\ln(\sqrt{\eta_\mathrm{S}} + \sqrt{\eta_\mathrm{S} - \epsilon}) \nonumber\\
    && + \sum_{j=1}^k \frac{1}{jC_{2j}^j}\frac{4^j}{\epsilon^j} \bigg(\eta_\mathrm{S}^{j} \sqrt{1 -  \frac{\epsilon}{\eta_\mathrm{S}}} + \eta_\mathrm{O}^{j} \sqrt{1 -  \frac{\epsilon}{\eta_\mathrm{O}}} \bigg)  \bigg] \nonumber\\
    && +\, \tilde{y}_{2k,m}^{+} \sum_{j=0}^k \frac{2C_k^j \epsilon^{k-j}}{2j+1} \bigg[ (\eta_\mathrm{O} - \epsilon)^{j+\frac{1}{2}} \nonumber\\
    && + (\eta_\mathrm{S} - \epsilon)^{j+\frac{1}{2}} \bigg] \bigg\}.
\end{eqnarray}

Similar to the discussion in Sec.~\ref{subsec_comparison_DeltaPhi}, the differences between our approximation \eqref{Ana_app_for_T_in_PS} and Eq.~\eqref{Ana_app_for_T_out_PS} appear in the logarithmic term and the sign of the last term, resulting in the inability to describe the travel times of photons in a unified way.

\subsubsection{Comparison with numerical results}

Although the travel time tends to diverge as the distance between the observer and the black hole increases as discussed before, the time difference between two photons can be convergent given that their travel times cancel each other out at large distances.
Therefore, the validity of the approximation \eqref{Ana_app_for_T_in_PS} can be verified by taking into account the time delay of two photons within the following scenario.

\begin{figure*}[t!]    
	\centering
	\includegraphics[width=1.0\textwidth]{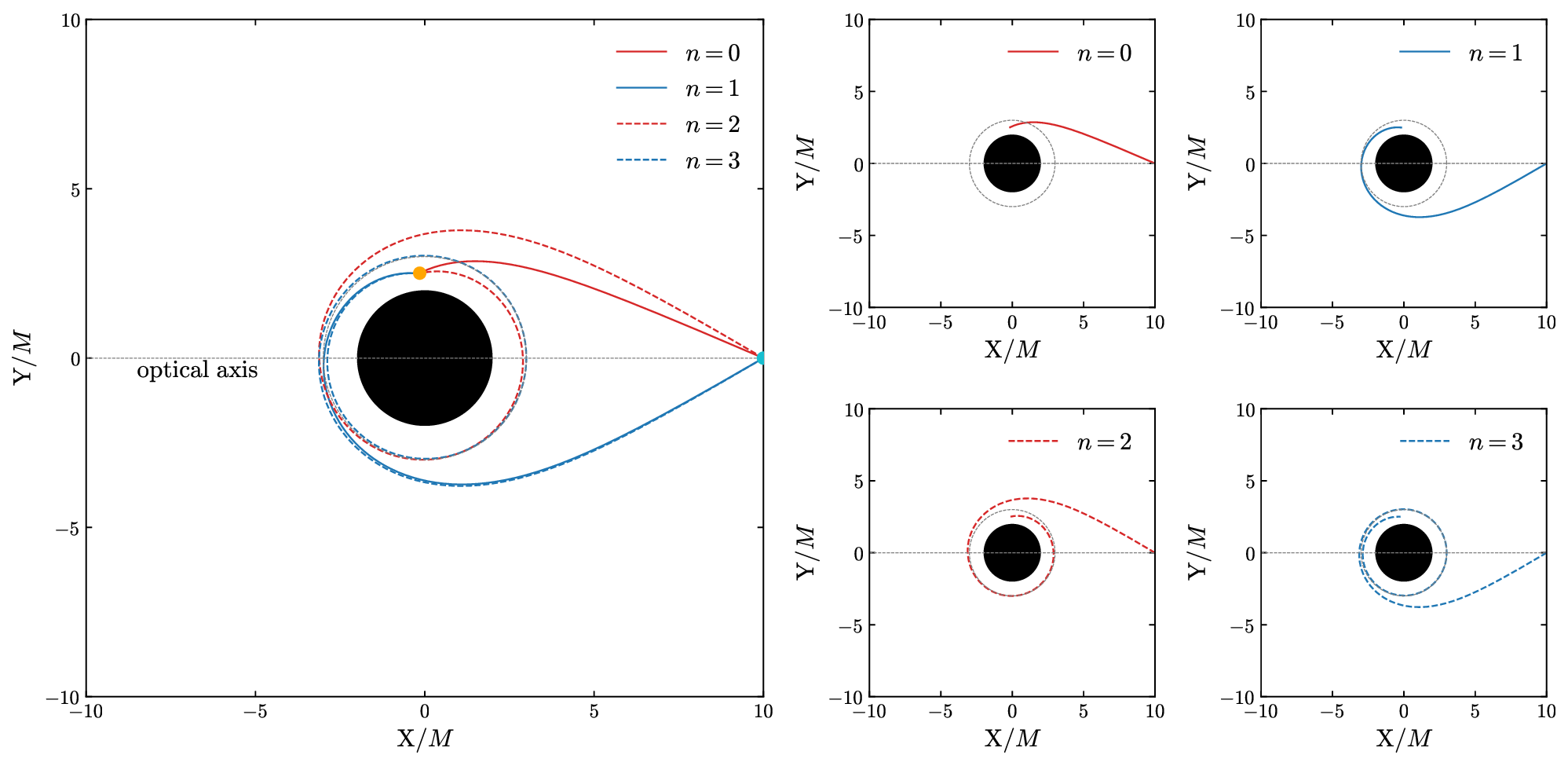}
	\caption{Different trajectories of photons emitted by a source inside the photon sphere that can be received by the observer. The central dark region is bounded by the black hole event horizon, and the dashed grey,  orange and sky blue circles represent the photon sphere, the source and the observer, respectively. The line connecting the observer and the black hole is the optical axis, and $n$ denotes the number of times that the photon intersects the optical axis before reaching the observer.}
	\label{Fig3_LightPath_in_BH_PS}
\end{figure*}

As shown in Fig.~\ref{Fig3_LightPath_in_BH_PS}, a source inside the photon sphere can emit different photons that can be received by the observer. 
These photons have different impact parameters, but their source has the same location, and so does the observer.
By defining the line connecting the observer and the black hole as the optical axis, we can classify the photons into types such as 
$n = 0, 1, 2, 3$ according to the number of times they cross the optical axis. 
This classification is very similar to the classification used when studying the photon rings generated by the accretion disk \cite{Gralla2019PRD100.024018,Tsupko2022PRD106.064033}. 
The photons with $n = 0$ originate from the direct emission of the source, while photons with $n = 1, 2, 3, \cdots$ are strongly deflected by the black hole.
They all obey the following lens equation 
\begin{eqnarray}\label{len_eq}
    \Delta \phi = \pi \pm ( \phi_\mathrm{O} - \phi_\mathrm{S} + \pi) + 2i\pi,\ \ i \geqslant 0,
\end{eqnarray} 
in which $\phi_\mathrm{S}$ and $\phi_\mathrm{O}$ are respectively the azimuthal angles of the source and the observer, and they are fixed for a given source and observer.
Eq.~\eqref{len_eq} is equivalent to Eq.~(36) in Ref.~\cite{Jia2021EPJC81.242}, and it reduces to the lens equation used in Refs.~\cite{Aratore2021JCAP10.054,Gao2024PRD109.063030} by setting $\phi_\mathrm{O} = \pi$.
In this work, we set $\phi_\mathrm{O} = 0$ without loss of generality.

Among the time delays between pairs of photons with $n = 0, 1, 2, 3, \cdots$, 
the time delay between photons with $n = 0$ and $n = 1$ is the most detectable since the observed intensity decreases exponentially as the crossing time $n$ increases \cite{Gralla2019PRD100.024018,Johnson2020SciAdv6.eaaz1310}.
By setting $i = 0$ in Eq.~\eqref{len_eq}, we can obtain the impact parameters $u|_{n=0}$ and $u|_{n=1}$
of photons with $n=0$ and $n=1$, repectively.
The time delay of these two photons can be defined as 
\begin{eqnarray}\label{Delta_T_in_PS}
    \Delta T & = & T( u|_{n=1}, r_\mathrm{S}, r_\mathrm{O} ) - T( u|_{n=0}, r_\mathrm{S}, r_\mathrm{O} ),
\end{eqnarray}
where the travel time $T$ is given by Eq.~\eqref{Ana_app_for_T_in_PS}.

By subsequently using Eqs.~\eqref{Ana_app_for_dphi_in_PS_order0}\eqref{Ana_app_for_T_in_PS_order0} and \eqref{relation_tf_and_bf}, at the lowest order $(N = 0)$ we can obtain
\begin{eqnarray}\label{Delta_T_in_PS_order0}
    \Delta T = 2(\pi - \phi_\mathrm{S}) u_\mathrm{m} + \mathrm{O}(\epsilon),
\end{eqnarray}
which can be used to roughly evaluate the magnitude of $\Delta T$. 
From Eq.~\eqref{Delta_T_in_PS_order0}, we can know that when  $\phi_\mathrm{S} = 0$, $\Delta T$ reaches its maximum value $2\pi u_\mathrm{m}$, which is approximately equal to the time taken for a photon to wind around the photon sphere once; when $\phi_\mathrm{S} = \pi$, we have $\Delta T = 0$, indicating that photons with $n=0$ and $n=1$ arrive at the observer simultaneously.

With the truncation order of $N=5$, we further use the time delay \eqref{Delta_T_in_PS} to test the accuracy of Eq.~\eqref{Ana_app_for_T_in_PS}.
The detailed procedure for calculating $\Delta T$ is given as follows.
\begin{itemize}
    \item[(1)] Given $r_\mathrm{S}, r_\mathrm{O}, \phi_\mathrm{S}$ and $\phi_\mathrm{O}$, calculate the impact parameters $u|_{n=0}$ and $u|_{n=1}$ of photons with $n = 0$ and $n = 1$ by solving the lens equation \eqref{len_eq} with $i = 0$. Here $\Delta \phi$ is given by Eq.~\eqref{Ana_app_for_dphi_in_PS} and a FindRoot algorithm is needed.
    \item[(2)] Calculate $T( u|_{n=0}, r_\mathrm{S}, r_\mathrm{O} )$ and $T( u|_{n=1}, r_\mathrm{S}, r_\mathrm{O} )$ by using Eq.~\eqref{Ana_app_for_T_in_PS}.
    \item[(3)] Compute $\Delta T$ via Eq.~\eqref{Delta_T_in_PS}.
\end{itemize}

\begin{figure}[htbp]    
	\centering
	\includegraphics[width=0.5\textwidth]{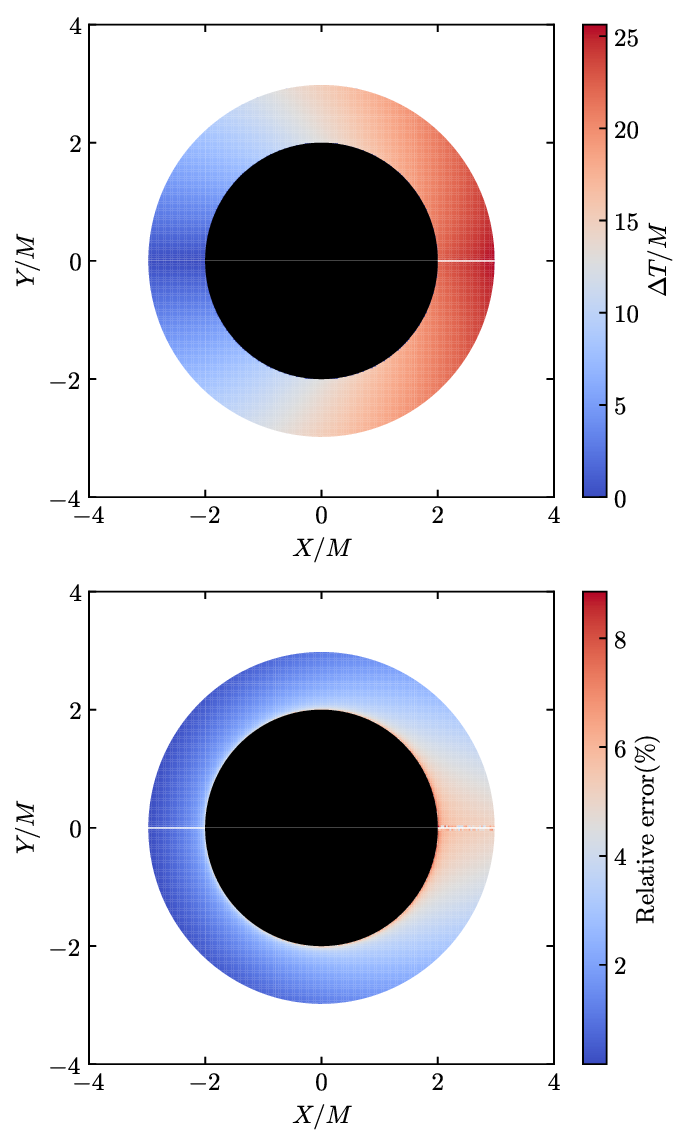}
	\caption{Top: The time delay $\Delta T$ calculated by Eq.~\eqref{Delta_T_in_PS}. The location of the observer is set to be $r_\mathrm{O} = 100M$ and $\phi_\mathrm{O} = 0$. Bottom: Relative errors between the analytical approximation \eqref{Delta_T_in_PS} and the numerical results. The order of approximation is taken to be $N = 5$. }
	\label{Fig4_RE_for_DeltaT}
\end{figure}

In the top panel of Fig.~\ref{Fig4_RE_for_DeltaT}, we show $\Delta T$ calculated by Eq.~\eqref{Delta_T_in_PS} with the truncation order being $N=5$. 
In the bottom panel, the relative errors of $\Delta T$ with respect to its numerical counterpart by evaluating $\big|( \Delta T_\mathrm{} - \Delta T_\mathrm{num} ) / \Delta T_\mathrm{num} \big|$ are illustrated.
The location of the observer is set to be $r_\mathrm{O} = 100M$.
In the numerical scheme, we obtain $u|_{n=0}$ and $u|_{n=1}$ that satisfy the lens equation \eqref{len_eq} by directly integrating the geodesic \eqref{geodesic}, and then calculate $\Delta T$ by integrating Eq.~\eqref{T_InBHPS}.

From the top panel of Fig.~\ref{Fig4_RE_for_DeltaT} we can observe that when $N=5$, the value of the time delay $\Delta T$ is consistent with $2(\pi - \phi_\mathrm{S}) u_\mathrm{m}$ predicted by Eq.~\eqref{Delta_T_in_PS_order0}. 
However, $\Delta T$ is not available for $\phi_\mathrm{S} = 0$.
This is because the impact parameter $u|_{n=0}$ is equal to 0 and its corresponding $\epsilon$ defined by \eqref{epsilon_def} becomes divergent, which causes Eq.~\eqref{Ana_app_for_T_in_PS} with $N=5$ to diverge. 
Such a problem always exists, but $\Delta T$ can be evaluated in the case of $N=0$ due to the relation \eqref{relation_tf_and_bf}. 
When $N=5$, the relation \eqref{relation_tf_and_bf} is corrected by higher-order terms, and one thus has to calculate $\Delta T$ from Eq.~\eqref{Ana_app_for_T_in_PS} directly. 

In the bottom panel of Fig.~\ref{Fig4_RE_for_DeltaT}, the relative errors of $\Delta T$ are not available in the case of $\phi_\mathrm{S} = 0$ and $\phi_\mathrm{S} = \pi$. 
The reasons are as follows. When $\phi_\mathrm{S} = 0$, Eq.~\eqref{Ana_app_for_T_in_PS} with $N=5$ is divergent as discussed above; when $\phi_\mathrm{S} = \pi$, the definition of the relative error is no longer valid since we have $\Delta T = 0$.
In other cases, the relative errors of $\Delta T$ are found to be well below 2\%, which proves the validity of the approximation \eqref{Ana_app_for_T_in_PS} indirectly.

\section{conclusions and discussion}
\label{sec_conclusion}

In this paper, the propagation of photons emitted by the source inside the photon sphere in the black hole spacetime is studied analytically.
To be specific, we present fully analytical approximations \eqref{Ana_app_for_dphi_in_PS} and \eqref{Ana_app_for_T_in_PS} for the change in the azimuthal angle and the travel time, respectively.
We also examine the accuracy of these formulas.

For photons with different impact parameters emitted by the source inside the photon sphere, their time delay can be quickly evaluated by employing our analytical formulas. 
We find that this delay is approximately $2(\pi - \phi_\mathrm{S}) u_\mathrm{m}$, where $\phi_\mathrm{S}$ is the azimuthal angle of the source, and $u_\mathrm{m}$ is the impact parameter at the photon sphere, which is related to the size of the black hole shadow \cite{Falcke2000ApJ528.L13}.

During the stage wherein the accretion flow enters the interior of the photon sphere, the aforementioned time delay may generate additional observational signatures in the received light curves, such as some peaks with specific time intervals \cite{CardenasAvendano2024PRL133.131402}, thus contributing to the precise determination of the shadow size. 
However, whether such an inflow event can be detected in the future needs further investigation.
Perhaps a three-dimensional numerical simulation is required \cite{Mummery2024MNRAS532.3395}.

Furthermore, when calculating the time delay, we have made the assumption that the source, lens and observer lie on a fixed plane, which is just a simple scenario. In more realistic contexts, for instance, when the source maintains a plunging orbit inside the photon sphere \cite{Mummery2022PRL129.161101}, the plane on which the source, lens and observer are located will change over time. How the resulting time delay evolves along the plunging orbit will be our next move.

Finally, black hole mimickers such as ultracompact objects can have an internal structure \cite{Cardoso2019LRR22.4,Virbhadra2002PRD65.103004,Tsukamoto2017PRD95.024030,Rosa2022PRD106.044031}, and photons emitted by the source inside their photon sphere may exhibit different time delay characteristics.
How to distinguish black holes and their mimickers through the time delay  also deserves future work.

\begin{acknowledgements}
We thank Prof. Yi Xie for valuable discussions.
This work is funded by the National Natural Science Foundation of China (Grants Nos. 12447143, 12273116, 62394350 and 62394351), the science research grants from the China Manned Space Project (Grants Nos. CMS-CSST-2021-A12 and CMS-CSST-2021-B10), the Strategic Priority Research Program of the Chinese Academy of Sciences (Grant No.XDA0350302) and the Opening Project of National Key Laboratory of Aerospace Flight Dynamics of China (Grant No. KGJ6142210220201).
\end{acknowledgements}

\appendix
\section{Derivation of the expression for $\bar{f}(\xi)$}
\label{App_Derivation_fxi}

The process of obtaining the series expansion of $\bar{f}(\xi)$ can be divided into the following four steps.
\begin{itemize}
    \item[(1)] According to the definition \eqref{e_and_xi_def} of $\xi$, we can obtain the series expansion of $\xi$ in terms of $( r - r_\mathrm{m} )$ as
    \begin{eqnarray}\label{xi_series}
        \xi = \sum_{i = 0}^\infty p_i ( r - r_\mathrm{m} )^i,
    \end{eqnarray}
    where $p_i$ is the coefficient of each term. Here the reason for the expansion around $r=r_\mathrm{m}$ is that photons are mainly deflected near the photon sphere. 
    Eq.~\eqref{xi_series} is applicable both to the case where the source is inside the photon sphere and to the case where the source is outside the photon sphere. 
    \item[(2)] Assuming that $q(\zeta)$ has the the form 
    \begin{eqnarray}\label{q_ansatz}
        q(\zeta) & = & \sum_{i = 0}^\infty q_i \xi^{\frac{i}{2}} \nonumber\\
        & = & \sum_{i = 0}^\infty q_i \bigg[ \sum_{j = 0}^\infty p_j ( r - r_\mathrm{m} )^j \bigg]^{\frac{i}{2}},
    \end{eqnarray}
    we can determine the coefficient $q_i$ by treating $( r - r_\mathrm{m} )$ as a small quantity and matching the coefficients of the following equation order-by-order
    \begin{eqnarray} \label{q_determination}
        \frac{1}{q} - r_\mathrm{m} = r - r_\mathrm{m}.
    \end{eqnarray}
    Furthermore, we can obtain the series expansion of $\mathrm{d}q / \mathrm{d}\zeta$ as 
    \begin{eqnarray}
        \frac{\mathrm{d}q}{\mathrm{d}\zeta} & = & \sum_{i = 0}^\infty \frac{i}{2} q_i \xi^{\frac{i}{2} - 1} \nonumber\\
        & = & \sum_{i = -1}^\infty Q_i \xi^{\frac{i}{2}}.
    \end{eqnarray}
    The assumption \eqref{q_ansatz} has been confirmed in the case where the source is outside the photon sphere \cite{Jia2021EPJC81.242}. We find it also holds for the case where the source is inside the photon sphere in a black hole spacetime. Moreover, we also find that the determination of $q_i$ needs to be handled separately for the two cases of $r < r_\mathrm{m}$ and $r > r_\mathrm{m}$. As a check, in the Schwarzschild black hole spacetime we have $(G = c = M = 1)$
    \begin{eqnarray}
        q = \frac{1}{3} + \frac{1}{3}\sqrt{\frac{2}{3}} \sqrt{\xi} - \frac{2}{27} \xi + \mathcal{O} \bigg(\xi^{\frac{3}{2}} \bigg)
    \end{eqnarray}
    for $r < r_\mathrm{m}$, while 
    \begin{eqnarray}
        q = \frac{1}{3} - \frac{1}{3}\sqrt{\frac{2}{3}} \sqrt{\xi} - \frac{2}{27} \xi + \mathcal{O} \bigg(\xi^{\frac{3}{2}} \bigg)
    \end{eqnarray}
    for $r > r_\mathrm{m}$.
    \item[(3)] Together with Eqs.~\eqref{q_def} and \eqref{q_determination}, we can derive the series expansion of $A(r), B(r), C(r)$ in terms of $\xi$. Then $\bar{f}(\xi)$ can be expressed as the product of the series expansions of all the terms.  
    \item[(4)] Collecting the powers of $\xi$, we can finally obtain Eq.~\eqref{fxi_for_DeltaPhi}.
\end{itemize}

\section{Integration formulas for $\epsilon<0$}
\label{App_integral_formulas}

By successively changing variables $\xi = \epsilon + s^2$ and $s = \sqrt{-\epsilon} \cosh(x)$, the integral $\int \frac{\xi^{\frac{n}{2}}}{\sqrt{\xi - \epsilon}}\mathrm{d}\xi$ becomes
\begin{eqnarray}
    \int_0^{\eta_i} \frac{\xi^{\frac{n}{2}}}{\sqrt{\xi - \epsilon}}\mathrm{d}\xi &=& 2(-\epsilon)^{\frac{n+1}{2}} \int_0^{x_i} \sinh^{n+1}(x) \mathrm{d}x.
\end{eqnarray}
For $n = 2k$, we have \cite{Gradshteyn2007Book}
\begin{eqnarray}
    \int_0^{\eta_i} \frac{\xi^{\frac{n}{2}}}{\sqrt{\xi - \epsilon}}\mathrm{d}\xi &=& 2(-\epsilon)^{k+\frac{1}{2}} \int_0^{x_i} \sinh^{2k+1}(x) \mathrm{d}x \nonumber\\
    &=& (-\epsilon)^{k+\frac{1}{2}} \sum_{j=0}^k \frac{2C_k^j(-1)^{k+j}}{2j+1} \nonumber\\
    && \times\bigg[ \bigg(\frac{\eta_i - \epsilon}{-\epsilon}\bigg)^{j+\frac{1}{2}} - 1 \bigg],
\end{eqnarray}
while for $n = 2k-1$, we have \cite{Gradshteyn2007Book}
\begin{eqnarray}
    \int_0^{\eta_i} \frac{\xi^{\frac{n}{2}}}{\sqrt{\xi - \epsilon}}\mathrm{d}\xi &=& 2(-\epsilon)^{k} \int_0^{x_i} \sinh^{2k}(x) \mathrm{d}x \nonumber\\
    &=& \frac{\epsilon^k C_{2k}^k}{4^k} \bigg[ -\ln(-\epsilon) + 2\ln(\sqrt{\eta_i} + \sqrt{\eta_i - \epsilon}) \nonumber\\
    && + \sum_{j=1}^k\frac{4^j}{jC_{2j}^j} \bigg(\frac{\eta_i}{\epsilon}\bigg)^{j} \sqrt{\frac{\eta_i-\epsilon}{\eta_i}} \bigg],
\end{eqnarray}
where $x_i = \mathrm{arccosh\bigg(\sqrt{\frac{\eta_i-\epsilon}{-\epsilon}}\bigg)}$.

\bibliographystyle{apsrev4-1}
\bibliography{Refs20230310} 

\end{document}